\documentclass[11pt,twoside]{article}
\usepackage[cp1251]{inputenc}  
\usepackage{graphicx}
\usepackage{amsmath}
\usepackage{amssymb}
\usepackage{mathrsfs}

\begin{document}

\thispagestyle{plain}

\newcommand{\bm}{\boldsymbol}
\newcommand{\mbf}{\mathbf}
\newcommand{\pst}{\hspace*{1.5em}}
\newcommand{\be}{\begin{equation}}
\newcommand{\ee}{\end{equation}}
\newcommand{\ds}{\displaystyle}
\newcommand{\bdm}{\begin{displaymath}}
\newcommand{\edm}{\end{displaymath}}
\newcommand{\bea}{\begin{eqnarray}}
\newcommand{\eea}{\end{eqnarray}}
\newcommand{\rmi}{\mathrm{i}}
\newcommand{\Tr}{\mathrm{Tr}}

\begin{center} {\Large \bf
\begin{tabular}{c}
Observables, evolution equation,\\
and stationary states equation in the joint probability \\
representation of quantum mechanics
\end{tabular}
 } \end{center}

\smallskip

\begin{center} {\bf Ya. A. Korennoy, V. I. Man'ko}\end{center}

\smallskip

\begin{center}
{\it P.N.    Lebedev Physical Institute,                          \\
       Leninskii prospect 53, 119991, Moscow, Russia }
\end{center}

\begin{abstract}\noindent
Symplectic and optical joint probability representations of quantum mechanics
are considered, in which the functions describing  the states are the probability distributions
with all random arguments (except the argument of time ). The general formalism 
of quantizers and dequantizers determining the star product
quantization scheme  in these representations is given. Taking the Gaussian functions 
as the distributions of the tomographic parameters 
the correspondence rules for most interesting physical operators are found
and the expressions of the dual symbols of operators in the form of singular 
and regular generalized functions are derived.
Evolution equations and stationary states equations
for symplectic and optical joint probability distributions are obtained.
\end{abstract}


\section{Introduction}

In Ref.~\cite{Mancini96} the probability representation of quantum
states was suggested (for a review see \cite{IbortPhysScr}). 
According to this representation the states of quantum systems are associated with fair 
probability distributions called quantum tomograms. 
The density operators of the quantum states can be determined from the tomograms,  and
consequently, the tomograms contain the complete information of the quantum properties
equivalent to the information embraced in all of the forms of the density operators
like Wigner function \cite{Wigner32}, Husimi function \cite{Husimi40}, 
Glauber-Sudarshan function \cite{Glauber63, Sudarshan63}.

Initially the quantum optical tomogram $w(X,\theta)$ was introduced as a tool 
for measuring the quantum state of radiation \cite{BerBer, VogRis}.
Generalizing the optical tomography technique the symplectic tomography
was introduced \cite{Mancini95}, and the evolution equation for symplectic tomograms
was found in \cite{Mancini96,ManciniFoundPhys97}.  
The problem of dual symbols of physical observables in the symplectic
tomography representation was considered in Ref.~\cite{OlgaJRLR97}.
Evolution equations for optical tomograms of spinless quantum systems
were obtained in Refs.~\cite{KorJRLR3274,KorJRLR32338}.
For the particles with spin it was done in~\cite{KorJRLR36534, KorIntJTP0163112}.
The correspondence rules and dual symbols of operators
in the optical tomography representation were obtained in \cite{KorPhysRevA85}.

The quantum state tomograms depend on extra parameters,
for example, the optical tomogram $w(X,\theta)$ \cite{BerBer, VogRis}
depend on the random position called quadrature component and the
parameter $\theta$ called local oscillator phase. 
It was pointed out 
\cite{BeautyInPhys, PhysScrT153} that the tomogram can be interpreted 
as a conditional probability distribution denoted as $w(X,\theta)\equiv w(X|\theta)$,
and such an interpretation provides the possibility to introduce the joint probability
distribution of two random variables $\widetilde w(X,\theta)$, which determines the optical tomogram
via Bayes' formula \cite{Bayes}.

The symplectic tomogram $M(X,\mu,\nu)$ \cite{Mancini95} represents the distribution function
of the position quadrature $X$ of rotated and squeezed (stretched) phase plane
determined by the parameters $\mu$ and $\nu$.
Thus, $M(X,\mu,\nu)\equiv  M(X|\mu,\nu)$ is a conditional distribution function of the variable 
$X$ under the condition of given $(\mu,\nu)$, and if the  distribution function for 
$\mu$ and $\nu$ is known, we can introduce the joint probability distribution 
$\widetilde M(X,\mu,\nu)$ of three random variables \cite{BeautyInPhys, PhysScrT153}.
Other tomographic schemes (like, e.g., spin tomography) also enable to construct
joint probability representations with all random variables (indices). 

The aim of this paper is derivation of correspondence rules and 
symbols of operators in the joint probability representation;
and foundation of evolution equations and stationary states equations for the joint probability
distributions. 

The paper is organized as follows. 
In Section 2 we introduce the  joint probability representation of states of quantum 
systems using the general formalism of quantizers and dequantizers.
In Section 3 we discuss correspondence rules for the operators, symbols of operators,
the evolution equation, and the equation of stationary states  in general case for arbitrary quantizer and
dequantizer.
In Section 4 we consider a specific example of symplectic joint probability representation 
in the $N-$dimensional case using shifted and scaled Gaussian distribution function 
for the tomographic parameters $\bm\mu$ and $\bm\nu$.
We find correspondence rules of operators, symbols of operators as singular and regular generalized
functions, evolution equation, and energy levels equation for joint probability distribution
of states. 
In Section 5 the optical joint probability representation with the 
distribution function for the phase vector $\bm\theta$ in the form of the weighted sum 
of  shifted and scaled Gaussian functions is presented.
Conclusions are given in Section 6.

\section{Joint probability representation of states of quantum systems}
Let  the density matrix $\hat\rho(t)$ dependent on time $t$ and normalized by the condition
$\mathrm{Tr}\{\hat\rho(t)\}=1$ corresponds to the state of the quantum system,
then in the tomographic representation this state is described by the tomographic distribution function
$\mathcal F(x,\eta,t)$ normalized by the condition
\be			\label{Normgeneral}
\int\mathcal F(x,\eta,t)dx=1,
\ee
where $x$  is a set of distribution variables and $\eta$ is a set of parameters 
of corresponding tomography.
According to the star product scheme (see \cite{SIGMA10086}), the tomogram is associated
with the density matrix in the following way:
\be			\label{DefTomgeneral}
\mathcal F(x,\eta,t)=\mathrm{Tr}\left\{
\hat\rho(t)\hat U_{\mathcal F}(x,\eta)
\right\},~~~
\hat\rho(t)=\int \hat D_{\mathcal F}(x,\eta)
\mathcal F(x,\eta,t) dxd\eta,
\ee
where $\hat U_{\mathcal F}(x,\eta)$ and $\hat D_{\mathcal F}(x,\eta)$ are
dequatizer and quantizer operators for appropriate tomographic scheme.

If we have spinless quantum system in the $N-$dimensional space,
then dequantizer and quantizer for the optical tomography \cite{KorPhysRevA85}
equal
\be		\label{dequantizerOPT}
\hat U_w(\mbf X,\bm\theta)=|\mbf X,\bm\theta\,\rangle\langle \mbf X,\bm\theta\,|=
\prod_{\sigma=1}^N
\delta\left(X_\sigma-\hat q_\sigma\cos\theta_\sigma-\hat p_\sigma\frac{\sin\theta_\sigma}
{m_\sigma\omega_{\sigma}}\right),
\ee
\be		\label{quantizerOPT}
\hat D_w(\mbf X,\bm\theta)=\int\prod_{\sigma=1}^N
\frac{\hbar\vert y_\sigma\vert}{2\pi m_\sigma\omega_{\sigma}}
\exp\left\{i y_\sigma\left(X_\sigma-\hat q_\sigma\cos\theta_\sigma
-\hat p_\sigma\frac{\sin\theta_\sigma}
{m_\sigma\omega_{\sigma}}\right)\right\}
d^Ny,
\ee
where $m_\sigma$ and $\omega_\sigma$ are constants that have dimensions
of mass and frequency and are chosen for reasons of convenience for the Hamiltonian 
of the quantum system under study,
$|\mbf X,\bm\theta\,\rangle$  is an eigenfunction of the operator
$\hat{\mbf X}(\bm\theta)$ with components
$\hat X_\sigma=\hat q_\sigma\cos\theta_\sigma+(\hat p_\sigma\sin\theta_\sigma)/(m_\sigma\omega_\sigma)$
corresponding to the eigenvalue $\mbf X$, where
$\hat{q}_\sigma$ and $\hat{p}_\sigma$ are the canonical 
position and momentum operators.

For the symplectic tomography the quantizer and dequantizer \cite{OVMankoJPhysA2002} can be written as:
\be		\label{dequantizerSYMP}
\hat U_M(\mbf X,\bm\mu,\bm\nu)=|\mbf X,\bm\mu,\bm\nu\,\rangle\langle \mbf X,\bm\mu,\bm\nu\,|=
\prod_{\sigma=1}^N
\delta(X_\sigma-\hat q_\sigma\mu_\sigma-\hat p_\sigma\nu_\sigma),
\ee
\be		\label{quantizerSYMP}
\hat D_M(\mbf X,\bm\mu,\bm\nu)=
\prod_{\sigma=1}^N\frac{m_\sigma\omega_{\sigma}}{2\pi}
\exp\left\{i\sqrt{\frac{m_\sigma\omega_{\sigma}}{\hbar}}
\left(X_\sigma-\hat q_\sigma\mu_\sigma-\hat p_\sigma\nu_\sigma\right)\right\},
\ee
where $|\mbf X,\bm\mu,\bm\nu\,\rangle$ is an eigenfunction of the operator
$\hat{\mbf X}(\bm\mu,\bm\nu)$ with components
$\hat X_\sigma=\mu_\sigma\hat q_\sigma+\nu_\sigma\hat p_\sigma$
corresponding to the eigenvalue $\mbf X$.

Symplectic and optical tomograms can also be found from the Wigner function
$W(\mbf q,\mbf p,t)$:
\be			\label{RelSympWig}
M(\mbf X,\bm\mu,\bm\nu,t)=\int\prod_{\sigma=1}^N
\delta(X_\sigma-\mu_\sigma q_\sigma-\nu_\sigma p_\sigma)W(\mbf q,\mbf p,t)
d^Nqd^Np,
\ee
\be			\label{RelOptWig}
w(\mbf X,\bm\theta,t)=\int\prod_{\sigma=1}^N
\delta\left(X_\sigma-q_\sigma\cos\theta_\sigma 
-p_\sigma\frac{\sin\theta_\sigma}{m_\sigma\omega_\sigma}\right)W(\mbf q,\mbf p,t)
d^Nqd^Np,
\ee
which, in turn, is determined by the density matrix in the position representation 
by the well-known formula
\be			\label{Def Wig}
W(\mbf q,\mbf p,t)=\frac{1}{(2\pi\hbar)^N}\int \rho(\mbf q+\mbf u/2,\mbf q-\mbf u/2,t)
e^{-i\mbf p\mbf u/\hbar}d^Nu.
\ee

Following by Refs. \cite{BeautyInPhys, PhysScrT153} let we introduce the joint probability
distribution function describing the state of a physical system.
For this aim we note that if the set of parameters $\eta$ is chosen randomly with a distribution function
$P(\eta)$ normalized by the condition
\be			\label{normir}
\int P(\eta)d\eta=1,
\ee
then, according to Bayes' formula \cite{Bayes}, dependent (generally speaking) on time
joint probability distribution $\widetilde{\mathcal F}(x,\eta,t)$ of the two sets 
of random variables $X$ and $\eta$ will be
equal
\be			\label{eq1}
\widetilde{\mathcal F}(x,\eta,t)= {\mathcal F}(x,\eta,t)P(\eta).
\ee

Due to normalization property of the tomogram (\ref{Normgeneral})
and normalized distribution function $P(\eta)$ of the set of parameters $\eta$,
the joint probability distribution (\ref{eq1}) will be automatically
normalized, but in the space of two sets of variables $x$ and $\eta$ 
\be			\label{eq2}
\int \widetilde {\mathcal F}(x,\eta,t) dxd\eta=1.
\ee
Further let us make use of the universal star product scheme (see, e.g. \cite{SIGMA10086}).
For this we introduce the corresponding dequantizer and quantizer operators relating the function
$\widetilde{\mathcal F}(x,\eta,t)$ and the density matrix $\hat\rho(t)$. It is evident that
\be			\label{eq3}
\widetilde{\mathcal F}(x,\eta,t) =\mathrm{Tr}\left\{
\hat\rho(t)\hat U_{\mathcal F}(x,\eta)P(\eta)
\right\}=
\mathrm{Tr}\left\{
\hat\rho(t) \hat U_{\widetilde{\mathcal F}}(x,\eta)
\right\},
\ee
\be			\label{eq4}
\hat\rho(t)=\int \hat D_{\mathcal F}(x,\eta)P^{-1}(\eta)
\widetilde{\mathcal F}(x,\eta,t) dxd\eta=
\int\hat D_{\widetilde{\mathcal F}}(x,\eta)
\widetilde{\mathcal F}(x,\eta,t) dxd\eta,
\ee
where $\hat U_{\widetilde{\mathcal F}}(x,\eta)$, $\hat D_{\widetilde{\mathcal F}}(x,\eta)$ are the new
dequantizer and quantizer for the star product scheme in the joint probability representation
\be			\label{eq5_1}
\hat U_{\widetilde{\mathcal F}}(x,\eta)=P(\eta) \hat U_{\mathcal F}(x,\eta),~~~~
\hat D_{\widetilde{\mathcal F}}(x,\eta)=P^{-1}(\eta) \hat D_{\mathcal F}(x,\eta).
\ee

Additionally, in the definition of the symplectic joint probability distribution we will assume 
that the distribution function $P(\bm\mu,\bm\nu)$ of  tomographic parameters tends to zero 
at infinity with all of its derivatives, and it is integrable across the hyperspace $(\bm\mu,\bm\nu)$
with any finite products of its arguments.

For the optical joint probability representation we note that the tomogram $w(\mbf x,\bm\theta,t)$
contains the whole of the available information about the state when all of the components of the phase
vector $\bm\theta$ are varied from zero to $\pi$, and it does not contain redundant information.

Therefore, in order that the joint distribution function $\widetilde w(\mbf x,\bm\theta)$ should also contain
all the information available on the state, the distribution function $P(\bm\theta)$ must not
turn to zero on the multitude $\left\{\theta_\sigma\in [0,\,\pi]\right\}$.
Besides, the function $P(\bm\theta)$ should be chosen so as to satisfy the normalization condition
$\int_0^\pi P(\bm\theta)d^N\theta=1$.

\section{Correspondence rules for the operators, evolution equation \\
and stationary states equation}

If an operator $\hat A$ defined on the set of density matrices
$\left\{\hat\rho\right\}$ acts on $\hat\rho$ as $\hat A\hat\rho$,
then, according to the general scheme, the action of this operator on 
$\widetilde{\mathcal F}(x,\eta)$ in the joint probability representation
can be expressed in terms of the operators $\hat U_{\widetilde{\mathcal F}}$ 
and $\hat D_{\widetilde{\mathcal F}}$ as follows (for brevity we shall omit the argument $t$,
assuming that the function $\widetilde {\mathcal F}(x,\eta)$ may depend on time):
\bea			
\big[\hat A\big]_{\widetilde{\mathcal F}} \widetilde{\mathcal F}(x,\eta)&=&
\mathrm{Tr}\left\{
\hat U_{\widetilde{\mathcal F}}(x,\eta)\hat A\int \hat D_{\widetilde{\mathcal F}}(x',\eta')
\widetilde{\mathcal F}(x',\eta')dx'd\eta'
\right\}, \nonumber \\[3mm]
&=&
\int\mathrm{Tr}\left\{
\hat U_{\widetilde{\mathcal F}}(x,\eta)\hat A \hat D_{\widetilde{\mathcal F}}(x',\eta')
\right\}\widetilde{\mathcal F}(x',\eta')dx'd\eta',
\label{eq6}
\eea
that is, in this representation the operator $\big[\hat A\big]_{\widetilde{\mathcal F}}$
is, generally speaking, an integral operator with the kernel
\be			\label{eq7}
\mathcal{K}(x,\eta,x',\eta')=\mathrm{Tr}\left\{
\hat U_{\widetilde{\mathcal F}}(x,\eta)\hat A \hat D_{\widetilde {\mathcal F}}(x',\eta')
\right\}.
\ee
With the help of formulae (\ref{eq6}) - (\ref{eq7}) and knowing the expression
for any operator in the density matrix representation
we can find its expression in the joint probability representation.
However, because of the simple relation between dequantizers and quantizers for the tomographic
representation and for the joint probability representation (\ref{eq5_1}),
the correspondence rules can be found directly from the appropriate rules in the relevant 
tomographic representation. That is, if $\big[\hat A\big]_{\mathcal F}$ is the expression
for the operator $\hat A$ in the tomographic representation, then in the joint probability
representation we, obviously, have:
\be			\label{eq8}
\big[\hat A\big]_{\widetilde {\mathcal F}}=P(\eta)\big[\hat A\big]_{\mathcal F}
P^{-1}(\eta).
\ee
Thus, for example, for the position operator
\be			\label{eq9}
[\hat{\mbf q}]_{\widetilde{\mathcal F}}=P(\eta)[\hat{\mbf q}]_{\mathcal F}
P^{-1}(\eta).
\ee
For the sum and for the product of two operators $\hat A$ and $\hat B$
we can write:
\bdm
\big[\hat A+\hat B\big]_{\widetilde{\mathcal F}}=\big[\hat A\big]_{\widetilde{\mathcal F}}
+\big[\hat B\big]_{\widetilde{\mathcal F}},~~~~
\big[\hat A\hat B\big]_{\widetilde{\mathcal F}}=\big[\hat A\big]_{\widetilde{\mathcal F}}
\big[\hat B\big]_{\widetilde{\mathcal F}}\,.
\edm
From these properties it follows that for any analytic function
$R(\hat A_1,\hat A_2,...,\hat A_k)$ on the set of operators
$\hat A_1,\hat A_2,...,\hat A_k$ the equality is fulfilled
\be			\label{ratifuncoper}
\big[R(\hat A_1,\hat A_2,...,\hat A_k)\big]_{\widetilde{\mathcal F}}=
R\left(\big[\hat A_1\big]_{\widetilde{\mathcal F}},\big[\hat A_2\big]_{\widetilde{\mathcal F}},...,
\big[\hat A_k\big]_{\widetilde{\mathcal F}}\right).
\ee
Thus, in most cases it is sufficient to find the correspondence rules
for position and momentum operators.

For any operator $\hat A$ its symbol $\widetilde{\mathcal F}_{\hat A}(x,\eta)$ and dual symbol
$\widetilde{\mathcal F}^{(d)}_{\hat A}(x,\eta)$ are also found
in accordance with the general scheme 
using dequantizer and quantizer (\ref{eq5_1})
\be			\label{symbgeneral0}
\widetilde{\mathcal F}_{\hat A}(x,\eta)=\mathrm{Tr}\left\{\hat A\hat U_{\widetilde{\mathcal F}}(x,\eta)\right\},
\ee
\be			\label{symbgeneral}
\widetilde{\mathcal F}^{(d)}_{\hat A}(x,\eta)=\mathrm{Tr}\left\{\hat A\hat D_{\widetilde{\mathcal F}}(x,\eta)\right\}.
\ee
The average value of the operator $\hat A$ in the state described by the joint probability distribution
$\widetilde{\mathcal F}(x,\eta)$ is determined by the dual symbol as follows:
\be			\label{averdual}
\langle\hat A\rangle=\int\widetilde{\mathcal F}^{(d)}_{\hat A}(x,\eta)
\widetilde{\mathcal F}(x,\eta)dx d\eta.
\ee

The evolution equation for the joint probability distribution is found from the von-Neumann equation
\be		\label{vonNeumann}
i\hbar\partial_t\hat\rho=[\hat H,\hat\rho]
\ee
according to the method \cite{KorJRLR32338}
\be			\label{evolvEQ1}
\partial_t\widetilde{\mathcal F}(x,\eta,t)=\frac{2}{\hbar}\int\mathrm{Im}\left[\mathrm{Tr}
\left\{\hat H(t)\hat D_{\widetilde{\mathcal F}}(x',\eta')\hat U_{\widetilde{\mathcal F}}(x,\eta)\right\}\right]\widetilde{\mathcal F}(x',\eta',t)
dx'd\eta',
\ee
or
\be			\label{evolvEQ2}
\partial_t \widetilde{\mathcal F}(x,\eta,t)=\frac{2}{\hbar}\mathrm{Im}\hat H
\left([\hat{\mbf q}]_{\widetilde{\mathcal F}}\,,[\hat{\mbf p}]_{\widetilde{\mathcal F}},t\right)
\widetilde{\mathcal F}(x,\eta,t).
\ee
where hereinafter $\partial_t $ is an abbreviated designation of the derivative $\partial/\partial t$.

When the Hamiltonian is time-independent, for the stationary states equation 
\be			\label{statEQ}
\hat H\hat\rho_E=E\hat\rho_E=\hat\rho_E\hat H
\ee
in the joint probability representation we have:
\be			\label{statEQ1}
E\widetilde{\mathcal F}_E(x,\eta)=\mathrm{Re}\hat H
\left([\hat{\mbf q}]_{\widetilde{\mathcal F}}\,,[\hat{\mbf p}]_{\widetilde{\mathcal F}}\right)
\widetilde{\mathcal F}_E(x,\eta).
\ee
Joint probability distributions $\widetilde{\mathcal F}_E(x,\eta)$ corresponding to the stationary states 
must also satisfy the stationary condition, which can be written as:
\be			\label{statCond}
\mathrm{Im}\hat H
\left\{[\hat{\mbf q}]_{\widetilde{\mathcal F}}\,,[\hat{\mbf p}]_{\widetilde{\mathcal F}}\right\}
\widetilde{\mathcal F}_E(x,\eta)=0.
\ee

\section{Symplectic joint probability representation with the Gaussian distribution
of tomographic parameters $\bm\mu$ and $\bm\nu$}

As an example, consider the case with the distribution of the tomographic parameters
$\bm\mu$ and $\bm\nu$ in the form of shifted and deformed Gaussian function
for $N-$dimensional quantum system
\be			\label{shiftGauss}
P_1(\bm{\mu},\bm{\nu})=\pi^{-N}\prod_{\sigma=1}^N\xi_\sigma^{-1} \zeta_\sigma^{-1}
\exp\left[-\frac{(\mu_\sigma-\mu_{0\sigma})^2}{\xi_\sigma^2}\right]
\exp\left[-\frac{(\nu_\sigma-\nu_{0\sigma})^2}{\zeta_\sigma^2}\right].
\ee
From the correspondence rules for the operators of components of position and momentum
in the symplectic tomography representation \cite{OlgaJRLR97}
\be			\label{corrulesQP}
[\hat q_j]_M=-\partial_{\mu_j}\partial^{-1}_{X_j}+
i\frac{\nu_j\hbar}{2}\partial_{X_j},~~~~
[\hat p_j]_M=-\partial_{\nu_j}\partial^{-1}_{X_j}-
i\frac{\mu_j\hbar}{2}\partial_{X_j},
\ee
with the help of formula (\ref{eq8})  we obtain the operators 
of position and momentum in the joint probability representation
\be			\label{QreprP2}
\left[\hat q_j\right]_{\widetilde M}=P_1(\bm{\mu},\bm{\nu})\left(-\partial_{X_j}^{-1}\partial_{\mu_j}
+i\frac{\nu_j\hbar}{2}\partial_{X_j}\right)P_1^{-1}(\bm{\mu},\bm{\nu})=
-\left(2\frac{\mu_j-\mu_{0j}}{\xi_j^2}+\partial_{\mu_j}\right)\partial_{X_j}^{-1}
+i\frac{\nu_j\hbar}{2}\partial_{X_j},
\ee
\be			\label{PreprP2}
\left[\hat p\right]_{\widetilde M}=
-\left(2\frac{\nu_j-\nu_{0j}}{\zeta_j^2}+\partial_{\nu_j}\right)\partial_{X_j}^{-1}
+i\frac{\mu_j\hbar}{2}\partial_{X_j}.
\ee
where we introduced the designation  for inverse derivatives \cite{KorPhysRevA85}
\be			\label{invder}
\partial_{x_\sigma}^{-n}F(x_\sigma)=\frac{1}{(n-1)!}
\int(x_\sigma-x_\sigma')^{n-1}\Theta(x_\sigma-x_\sigma')F(x_\sigma')d x_\sigma',
\ee
where $\Theta(x_\sigma-x_\sigma')$ is a Heaviside step function.

For the creation $\hat a_j$ and annihilation $\hat a^\dagger_j$ operators in the 
symplectic tomography representation we know \cite{OlgaJRLR97} that
\bdm
[\hat a_j]_M=\sqrt{\frac{m_j\omega_j}{2\hbar}}\left\{\frac{\hbar}{2}\partial_{X_j}\left(\frac{\mu_j}{m_j\omega_j}+i\nu_j\right)
-\partial^{-1}_{X_j}\left(\partial_{\mu_j}+\frac{i\partial_{\nu_j}}{m_j\omega_j}\right)\right\},
\edm
\bdm
[\hat a_j^\dagger]_M=\sqrt{\frac{m_j\omega_j}{2\hbar}}\left\{\frac{\hbar}{2}\partial_{X_j}
\left(\frac{-\mu_j}{m_j\omega_j}+i\nu_j\right)
-\partial^{-1}_{X_j}\left(\partial_{\mu_j}-\frac{i\partial_{\nu_j}}{m_j\omega_j}\right)\right\}.
\edm
Consequently, in accordance with (\ref{eq8}), in the joint probability representation we have:
\be			\label{AreprP2}
[\hat a_j]_{\widetilde M}=\sqrt{\frac{m_j\omega_j}{2\hbar}}\left\{
\frac{\hbar}{2}\partial_{X_j}\left(\frac{\mu_j}{m_j\omega_j}+i\nu_j\right)-\partial_{X_j}^{-1}
\left(\partial_{\mu_j}+i\frac{\partial_{\nu_j}}{m_j\omega_j}+2\frac{\mu_j-\mu_{0j}}{\xi_j^2} 
+i2\frac{\nu_j-\nu_{0j}}{m_j\omega_j\zeta_j^2}
\right)
\right\},
\ee
\be			\label{ATreprP2}
[\hat a^\dag_j]_{\widetilde M}=\sqrt{\frac{m_j\omega_j}{2\hbar}}\left\{
\frac{\hbar}{2}\partial_{X_j}\left(\frac{-\mu_j}{m_j\omega_j}+i\nu_j\right)-\partial_{X_j}^{-1}
\left(\partial_{\mu_j}-i\frac{\partial_{\nu_j}}{m_j\omega_j}+2\frac{\mu_j-\mu_{0j}}{\xi_j^2} 
-i2\frac{\nu_j-\nu_{0j}}{m_j\omega_j\zeta_j^2}
\right)
\right\}.
\ee
It is easy to check that the equality 
$\left[[\hat a_j]_{\widetilde M},[\hat a_j^\dagger]_{\widetilde M}\right]=1$ is performed,
because at the transition to the joint probability representation the
commutation relations are maintained.

Using the general definition (\ref{symbgeneral}) one can obtain the dual symbols of any operators.
Thus, for the identity operator $\hat1$, for the components $\hat q_j$ and $\hat p_j$,
and for the product $\hat q_j\hat p_j$ after some calculations we can write:
\be			\label{NDyalsymb1P2}
\widetilde M_{\hat 1}^{(d)}(\mbf{X},\bm{\mu},\bm{\nu})=
\pi^N\delta(\bm\mu)\delta(\bm\nu)\prod_\sigma
\xi_\sigma \zeta_\sigma\exp\left(\frac{\mu_{0\sigma}^2}{\nu_{0\sigma}^2}
+\frac{\nu_{0\sigma}^2}{\zeta_\sigma^2}\right),
\ee
\bdm
\widetilde M^{(d)}_{\hat q_j}(\mbf X,\bm\mu,\bm\nu)=i\frac{\pi^N\sqrt\hbar}{\sqrt{m_j\omega_j}}
\left[2\frac{\mu_{0j}}{\xi_j^2}+\partial_{\mu_j}\right]\delta(\bm\mu)\delta(\bm\nu)
\prod_{\sigma=1}^N
\xi_\sigma\zeta_\sigma \exp\left(\frac{\mu_{0\sigma}^2}{\xi_\sigma^2}
+\frac{\nu_{0\sigma}^2}{\zeta_\sigma^2}+
i\sqrt{\frac{m_\sigma\omega_\sigma}{\hbar}}X_\sigma\right),
\edm 
\bdm
\widetilde M^{(d)}_{\hat p_j}(\mbf X,\bm\mu,\bm\nu)=i\frac{\pi^N\sqrt\hbar}{\sqrt{m_j\omega_j}}
\left[2\frac{\nu_{0j}}{\zeta_j^2}+\partial_{\nu_j}\right]\delta(\bm\mu)\delta(\bm\nu)
\prod_{\sigma=1}^N
\xi_\sigma\zeta_\sigma \exp\left(\frac{\mu_{0\sigma}^2}{\xi_\sigma^2}
+\frac{\nu_{0\sigma}^2}{\zeta_\sigma^2}+
i\sqrt{\frac{m_\sigma\omega_\sigma}{\hbar}}X_\sigma\right),
\edm 
\bea
\widetilde M^{(d)}_{\hat q_j\hat p_j}(\mbf X,\bm\mu,\bm\nu)&=&\pi^N\left\{\frac{-\hbar}{m_j\omega_j}\left[
\frac{2\mu_{0j}}{\xi_j^2}\delta(\bm\mu)+\delta'_{\mu_j}(\bm\mu)\right]
\left[\frac{2\nu_{0j}}{\zeta_j^2}\delta(\bm\nu)+\delta'_{\nu_j}(\bm\nu)\right]
+\frac{i\hbar}{2}\delta(\bm\mu)\delta(\bm\nu)\right\} \nonumber \\
&&\times\prod_{\sigma=1}^N
\xi_\sigma\zeta_\sigma \exp\left(\frac{\mu_{0\sigma}^2}{\xi_\sigma^2}
+\frac{\nu_{0\sigma}^2}{\zeta_\sigma^2}+
i\sqrt{\frac{m_\sigma\omega_\sigma}{\hbar}}X_\sigma\right).
\eea 

The dual symbol $\widetilde M_{\hat A}^{(d)}(\mbf X,\bm\mu,\bm\nu)$
of some operator $\hat A$ according to formula (\ref{averdual}) defines a linear
continuous functional on the multitude of joint distribution functions
$\widetilde M(\mbf X,\bm\mu,\bm\nu)$, that is, the set of $\widetilde M_{\hat A}^{(d)}(\mbf X,\bm\mu,\bm\nu)$
actually specifies a set of generalized functions on the multitude
$\left\{\widetilde M(\mbf X,\bm\mu,\bm\nu)\right\}$.
It is clear that the equality of two different symbols of one operator must be treated as
the equality of the generalized functions, i.e., two symbols are equal each other if and only if for any
joint distribution function we have the equality of the values of the functionals defined by these symbols.
In that way, generally speaking, for any operator $\hat A$ a set of symbols exist, which are equal
in the sense of generalized functions (\ref{averdual}).

Therefore, the dual symbols listed above are not uniquely defined, and we can write the other symbols
of the same operators, for example, the dual symbol of the component of position operator from the physical meaning of the 
symplectic tomogram up to the normalization factor equals
\be			\label{DyalsymbQP2}
\widetilde M_{\hat q_j}^{(d)}(\mbf{X},\bm{\mu},\bm{\nu})\sim 
X_j\delta(\mu_j-\xi_j)\delta(\bm\nu)\prod_{\sigma\not= j}\delta(\mu_\sigma).
\ee
The normalization factor is found from the equality
\bdm
\langle\hat q_j\rangle=\int q_jW(\mbf q,\mbf p,t)d^Nqd^Np=
\int\widetilde M_{\hat q_j}^{(d)}(\mbf{X},\bm{\mu},\bm{\nu})P_1(\bm{\mu},\bm{\nu})
M(\mbf{X},\bm{\mu},\bm{\nu})d^NXd^N\mu d^N\nu.
\edm
After calculations we obtain the final result
\be			\label{NDyalsymbQP2}
\widetilde M_{\hat q_j}^{(d)}(\mbf{X},\bm{\mu},\bm{\nu})=
\pi^N\zeta_j\exp\left(\frac{(\xi_j-\mu_{0j})^2}{\xi_j^2}+\frac{\nu_{0j}^2}{\zeta_j^2}\right)
X_j\delta(\mu_j-\xi_j)\delta(\bm\nu)\prod_{\sigma\not= j}
\xi_\sigma \zeta_\sigma\exp\left(\frac{\mu_{0\sigma}^2}{\xi_\sigma^2}+\frac{\nu_{0\sigma}^2}{\zeta_\sigma^2}\right)
\delta(\mu_\sigma).
\ee
Similarly we find dual symbols of the components of momentum $\hat p_j$, of the product of components
$\hat q_j\hat p_j$, and of the powers of components $\hat q_j^n$, $\hat p_j^n$:
\be			\label{NDyalsymbPP2}
\widetilde M_{\hat p_j}^{(d)}(\mbf{X},\bm{\mu},\bm{\nu})=
\pi^N\xi_j\exp\left(\frac{\mu_{0j}^2}{\xi_j^2}+\frac{(\zeta_j-\nu_{0j})^2}{\zeta_j^2}\right)
X_j\delta(\bm\mu)\delta(\nu_j-\zeta_j)\prod_{\sigma\not= j}
\xi_\sigma \zeta_\sigma\exp\left(\frac{\mu_{0\sigma}^2}{\xi_\sigma^2}+\frac{\nu_{0\sigma}^2}{\zeta_\sigma^2}\right)
\delta(\nu_\sigma),
\ee
\bea			\label{NDyalsymbQPP2}
&&\widetilde M_{\hat q_j\hat p_j}^{(d)}(\mbf{X},\bm{\mu},\bm{\nu})=
\frac{\pi^N}{2}X_j^2
\prod_{\sigma\not= j}
\xi_\sigma \zeta_\sigma\exp\left(\frac{\mu_{0\sigma}^2}{\xi_\sigma^2}+\frac{\nu_{0\sigma}^2}{\zeta_\sigma^2}\right)
\delta(\mu_\sigma)\delta(\nu_\sigma) \nonumber \\
&&\times\Bigg\{\delta(\mu_j-\xi_j)\delta(\nu_j-\zeta_j)
\exp\left(\frac{(\xi_j-\mu_{0j})^2}{\xi_j^2}+\frac{(\zeta_j-\nu_{0j})^2}{\zeta_j^2}\right)
-\delta(\mu_j-\xi_j)\delta(\nu_j)
\exp\left(\frac{(\xi_j-\mu_{0j})^2}{\xi_j^2}+\frac{\nu_{0j}^2}{\zeta_j^2}\right) \nonumber \\
&&-\delta(\mu_j)\delta(\nu_j-\zeta_j)
\exp\left(\frac{\mu_{0j}^2}{\xi_j^2}+\frac{(\zeta_j-\nu_{0j})^2}{\zeta_j^2}\right)
\Bigg\}+\frac{i\pi^N}{2}\delta(\bm\mu)\delta(\bm\nu)\prod_\sigma
\xi_\sigma \zeta_\sigma\exp\left(\frac{\mu_{0\sigma}^2}{\xi_\sigma^2}
+\frac{\nu_{0\sigma}^2}{\zeta_\sigma^2}\right),
\eea
\be			\label{NDyalsymbQ2P2}
\widetilde M_{\hat q_j^n}^{(d)}(\mbf{X},\bm{\mu},\bm{\nu})=
\pi^N\frac{\zeta_j}{\xi_j^{n-1}}\exp\left(\frac{(\xi_j-\mu_{0j})^2}{\xi_j^2}+\frac{\nu_{0j}^2}{\zeta_j^2}\right)
X_j^n\delta(\mu_j-\xi_j)\delta(\bm\nu)\prod_{\sigma\not= j}
\xi_\sigma \zeta_\sigma\exp\left(\frac{\mu_{0\sigma}^2}{\xi_\sigma^2}+\frac{\nu_{0\sigma}^2}{\zeta_\sigma^2}\right)
\delta(\mu_\sigma),
\ee
\be			\label{NDyalsymbP2P2}
\widetilde M_{\hat p_j^n}^{(d)}(\mbf{X},\bm{\mu},\bm{\nu})=
\pi^N\frac{\xi_j}{\zeta_j^{n-1}}\exp\left(\frac{\mu_{0j}^2}{\xi_j^2}+\frac{(\zeta_j-\nu_{0j})^2}{\zeta_j^2}\right)
X_j^n\delta(\bm\mu)\delta(\nu_j-\zeta_j)\prod_{\sigma\not= j}
\xi_\sigma \zeta_\sigma\exp\left(\frac{\mu_{0\sigma}^2}{\xi_\sigma^2}+\frac{\nu_{0\sigma}^2}{\zeta_\sigma^2}\right)
\delta(\nu_\sigma),
\ee
The dual symbol of the number of photons operator $\widetilde M_{\hat a_j^\dag\hat a_j}^{(d)}(\mbf{X},\bm{\mu},\bm{\nu})$
is expressed from the found symbols as follows:
\bdm
\widetilde M_{\hat a_j^\dag\hat a_j}(\mbf{X},\bm{\mu},\bm{\nu})=\frac{1}{2}\left(
\frac{m_j\omega_j}{\hbar}\widetilde M_{\hat q_j^2}^{(d)}(\mbf{X},\bm{\mu},\bm{\nu})+
\frac{1}{\hbar m_j\omega_j}\widetilde M_{\hat p_j^2}^{(d)}(\mbf{X},\bm{\mu},\bm{\nu})-
\widetilde M_{\hat 1}^{(d)}(\mbf{X},\bm{\mu},\bm{\nu})
\right).
\edm

Thus, we obtained appearances of  the dual symbols in the form of singular generalized functions.
But in the joint probability representation as in the optical tomography representation
\cite{KorPhysRevA85} the dual symbols of the operators can be expressed in the form of 
regular generalized functions.
To calculate such symbols we preliminary note that according to (\ref{RelSympWig})
\be			\label{prodXever}
\int\left[\prod_{\sigma=1}^N (X_\sigma)^{\alpha_\sigma}\right]M(\mbf X,\bm\mu,\bm\nu)d^NX=
\int\left[\prod_{\sigma=1}^N (\mu_\sigma q_\sigma+\nu_\sigma p_\sigma)^{\alpha_\sigma}\right]
W(\mbf q,\mbf p)d^Nqd^Np.
\ee
Next, we consider the integral of the following form
\be			\label{Zeroint}
\mathcal{I}=\int \mu_1^{\alpha_1}...\mu_N^{\alpha_N}
\nu_1^{\beta_1}...\nu_N^{\beta_N}
\frac{\partial^{k_1+...+k_N+l_1+...+l_N}P(\bm\mu,\bm\nu)}
{\partial\mu_1^{k_1}...\partial\mu_N^{k_N}\partial\nu_1^{l_1}...\partial\nu_N^{l_N}}
d^N\mu\, d^N\nu.
\ee
If at least one $k_j>\alpha_j$ or $l_j>\beta_j$, then this integral is equal to zero.
If $\alpha_j=k_j$ and $\beta_j=l_j$ for all $j=1,...N$, then
\be			\label{Zeroint1}
\mathcal{I}=\prod_{\sigma=1}^N(-1)^{k_\sigma+l_\sigma}k_\sigma!\,l_\sigma!\,\,\,.
\ee
Taking into account (\ref{prodXever}) and the specified properties of the integral (\ref{Zeroint})
we can write
\bea			\label{QPwignerever}
&&\!\!\!\!\!\!\!\!\!\!\!\!\!\!\!\!\!\!
\int\left[\prod_{\sigma=1}^Nq_\sigma^{k_\sigma}p_\sigma^{l_\sigma}\right]
W(\mbf q,\mbf p)d^Nqd^Np= \nonumber \\
&&\int\left[\prod_{\sigma=1}^N(-1)^{k_\sigma+l_\sigma}
\frac{X_\sigma^{(k_\sigma+l_\sigma)}}{(k_\sigma+l_\sigma)!}
\right]\frac{\partial^{k_1+...+k_N+l_1+...+l_N}P(\bm\mu,\bm\nu)}
{\partial\mu_1^{k_1}...\partial\mu_N^{k_N}\partial\nu_1^{l_1}...\partial\nu_N^{l_N}}
\frac{\widetilde M(\mbf X,\bm\mu,\bm\nu)}{P(\bm\mu,\bm\nu)}
d^NX d^N\mu\, d^N\nu.
\eea

Recall that for any operator $\hat A$ acting on the density matrix
you can find the appropriate operator in the Wigner-Weyl representation $\big[\hat A\big]_W$
acting on the Wigner function. It is well known that
(see,e.g., \cite{IbortPhysScr})
\bdm
\hat q_j\hat\rho~\leftrightarrow~ [\hat q_j]_W W(\mbf q,\mbf p)=
\left(q_j+\frac{i\hbar}{2}\partial_{p_j}\right)W(\mbf q,\mbf p),
\edm
\bdm
\hat p_j\hat\rho~\leftrightarrow~ [\hat p_j]_W W(\mbf q,\mbf p)=
\left(p_j-\frac{i\hbar}{2}\partial_{q_j}\right)W(\mbf q,\mbf p).
\edm
Therefore, the average value of any combination of the components of operators 
$\hat{\mbf q}$ and $\hat{\mbf p}$ can be expressed through a combination of integrals
(\ref{QPwignerever}), and thus, we can find a regular symbol of any operator interesting to us.
For example, let's find the regular symbol of the component $\hat q_j$,
\bea			\label{chainforQ}
\langle\hat q_j\rangle&=&\int\widetilde M_{\hat q_j}^{(d)}(\mbf X,\bm\mu,\bm\nu)
\widetilde M(\mbf X,\bm\mu,\bm\nu)d^NX d^N\mu\,d^N\nu=\int q_jW(\mbf q,\mbf p)d^Nqd^Np
\nonumber \\
&&=
-\int X_j\frac{\partial P(\bm\mu,\bm\nu)}{\partial\mu_j}
\frac{\widetilde M(\mbf X,\bm\mu,\bm\nu)}{P(\bm\mu,\bm\nu)}d^NX d^N\mu\,d^N\nu.
\eea
Thus, we have
\be			\label{Qsymbdualgen}
\widetilde M_{\hat q_j}^{(d)}(\mbf X,\bm\mu,\bm\nu)=
-\frac{X_j}{P(\bm\mu,\bm\nu)}\frac{\partial P(\bm\mu,\bm\nu)}{\partial\mu_j}.
\ee
Substituting (\ref{shiftGauss}) instead of arbitrary distribution $P(\bm\mu,\bm\nu)$ we obtain
\be			\label{regDyalQP2}
\widetilde M_{\hat q_j}^{(d)}(\mbf{X},\bm{\mu},\bm{\nu})=
2\frac{\mu_j-\mu_{0j}}{\xi_j^2}X_j.
\ee
Direct verification shows that
\bdm
\langle\hat q_j\rangle=\int2\frac{\mu_j-\mu_{0j}}{\xi_j^2}X_j 
P_1(\bm{\mu},\bm{\nu})
M(\mbf{X},\bm{\mu},\bm{\nu})d^NXd^N\mu d^N\nu=
\int q_jW(\mbf q,\mbf p)d^Nqd^Np.
\edm
Taking into account the symmetry considerations between the operators 
$\hat{\mbf q}$, $\hat{\mbf p}$ in the definition of the symplectic tomogram
and the symmetry of the function $P_1(\bm{\mu},\bm{\nu})$ 
from the symbol for position (\ref{regDyalQP2}) we can obtain the dual symbol
for the component of momentum
\be			\label{regDyalPP2}
\widetilde M_{\hat p_j}^{(d)}(\mbf{X},\bm{\mu},\bm{\nu})=
2\frac{\nu_j-\nu_{0j}}{\zeta_j^2}X_j.
\ee
Similarly dual symbols for other operators in the form of regular generalized functions are calculated:
\be			\label{regDyalQ2P2}
\widetilde M_{\hat q_j^2}^{(d)}(\mbf{X},\bm{\mu},\bm{\nu})=
\frac{X_j^2}{\xi_j^4}\left[2(\mu_j-\mu_{0j})^2-\xi_j^2\right],
\ee
\be			\label{regDyalP2P2}
\widetilde M_{\hat p_j^2}^{(d)}(\mbf{X},\bm{\mu},\bm{\nu})=
\frac{X_j^2}{\zeta_j^4}\left[2(\nu_j-\nu_{0j})^2-\zeta_j^2\right],
\ee
\be			\label{regDyalQPP2}
\widetilde M_{\hat q_j\hat p_j}^{(d)}(\mbf{X},\bm{\mu},\bm{\nu})=
2X_j^2\,\frac{\mu_j-\mu_{0j}}{\xi_j^2}\,\frac{\nu_j-\nu_{0j}}{\zeta_j^2}+\frac{i\hbar}{2},
\ee
\be			\label{regDyalaa2}
\widetilde M_{\hat a_j^\dag\hat a_j}^{(d)}(\mbf{X},\bm{\mu},\bm{\nu})=
\frac{X_j^2m_j\omega_j}{2\hbar}\left[\frac{2(\mu_j-\mu_{0j})^2-\xi_j^2}{\xi_j^4}
+\frac{2(\nu_j-\nu_{0j})^2-\zeta_j^2}{m_j^2\omega_j^2\zeta_j^4}
\right]-\frac{1}{2}.
\ee
Dual symbols in the form of regular generalized functions  are also  defined non-uniquely, but all of their versions
for the same operators are equal each other in the sense of generalized functions.
E.g., the immediate verification shows that the symbols
\bdm
\widetilde M^{(d)}_{\hat q_j^2}=\frac{X_j^2}{2\xi_j^2}\left(\frac{3\mu_j^2}{\xi_j^2}-\frac{\nu_j^2}{\zeta_j^2}\right)
\exp\left(-\mu_{0j}\frac{2\mu_j-\mu_{0j}}{\xi_j^2}
-\nu_{0j}\frac{2\nu_j-\nu_{0j}}{\zeta_j^2} \right),
\edm
\bdm
\widetilde M^{(d)}_{\hat p_j^2}=\frac{X_j^2}{2\zeta_j^2}\left(\frac{3\nu_j^2}{\zeta_j^2}-\frac{\mu_j^2}{\xi_j^2}\right)
\exp\left(-\mu_{0j}\frac{2\mu_j-\mu_{0j}}{\xi_j^2}
-\nu_{0j}\frac{2\nu_j-\nu_{0j}}{\zeta_j^2} \right),
\edm
give rise to the correct average values $\langle q_j^2\rangle$ and $\langle p_j^2\rangle$
respectively.

Further, let's write the evolution equation of the joint probability distribution.
For this we use the previously found correspondence rules (\ref{QreprP2}) and (\ref{PreprP2})
for the components of the operators $\hat{\mbf q}$ and $\hat{\mbf p}$.
From formula (\ref{evolvEQ2}) with the Hamiltonian of the form
\be			\label{eq36pr}
\hat H=\sum_{\sigma=1}^N\frac{\hat p_\sigma^2}{2m_\sigma}+V(\mbf q,t)
\ee
after some calculations we obtain:

\be			\label{evolvEQP2}
\partial_t\widetilde M(\mbf{X},\bm{\mu},\bm{\nu},t)=
\left[
\sum_{j=1}^N \frac{\mu_j}{m_j}\left(2\frac{\nu_j-\nu_{0j}}{\zeta_j^2}+\partial_{\nu_j}\right)
+\frac{2}{\hbar}\mathrm{Im}V\Big([\hat{\mbf q}]_{\widetilde M},t\Big)
\right]
\widetilde M(\mbf{X},\bm{\mu},\bm{\nu},t).
\ee

The equation for the joint distributions of stationary states,
when the potential $V$ is time-independent, is found using (\ref{statEQ1})
\bea			\label{statEQP2}
&&\!\!\!\!\!\!\!\!\!\!\!\!
\left[E-\mathrm{Re}V\Big([\hat{\mbf q}]_{\widetilde M}\Big)\right]\widetilde M_E(\mbf{X},\bm{\mu},\bm{\nu})=
\nonumber \\
&&=\sum_{j=1}^N\left[\frac{\partial_{X_j}^{-2}}{m_j}\left(2\frac{(\nu_j+\nu_{0j})^2}{\zeta_j^4}+\frac{\partial_{\nu_j}^2}{2}
+2\frac{\nu_j+\nu_{0j}}{\zeta_j^2}\partial_{\nu_j}+\frac{1}{\zeta_j^2}
\right)-\frac{\mu_j^2\hbar^2}{8m_j}\partial_{X_j}^2\right]
\widetilde M_E(\mbf{X},\bm{\mu},\bm{\nu}).
\eea
The distribution functions $\widetilde M_E(\mbf{X},\bm{\mu},\bm{\nu})$ also must satisfy 
by the stationary condition
\be			\label{statCondP2}
\left[
\sum_{j=1}^N \frac{\mu_j}{m_j}\left(\frac{\nu_j-\nu_{0j}}{\zeta_j^2}+\frac{\partial_{\nu_j}}{2}\right)
+\frac{1}{\hbar}\mathrm{Im}V\Big([\hat{\mbf q}]_{\widetilde M}\Big)
\right]
\widetilde M_E(\mbf{X},\bm{\mu},\bm{\nu})=0,
\ee
which is obtained from equation (\ref{evolvEQP2}) at $\partial_t\widetilde M_E(\mbf{X},\bm{\mu},\bm{\nu})=0$.

\section{Optical joint probability representation with the sum of Gaussian distributions for phase vector $\bm\theta$}
Next, consider the representation of the joint distribution function $\widetilde w(\mbf X,\bm\theta)$ 
for $N-$dimensional quantum system, in which the distribution of the parameters  $\bm\theta$ is assumed 
to be the following weighted sum:
\be			\label{NshiftedGauss}
P_2(\bm{\theta})=\sum_{k=1}^KQ_k{\mathcal P}_k(\bm{\theta}),
~~~~
\sum_{k=1}^K Q_k=1.
\ee
where $\{Q_k\}$ is a set of weights, $\{{\mathcal P}_k(\bm{\theta})\}$ is a set of 
$N-$dimensional Gaussians
\bdm
{\mathcal P}_k(\bm{\theta})={\mathcal N}_k\prod_{\sigma=1}^N
\exp\left[-\frac{(\theta_\sigma-f_{k\sigma})^2}{\phi_{k\sigma}^2}\right]
\edm
with the normalization factors $\{{\mathcal N}_k\}$ such that
\bdm
\int\limits_0^\pi{\mathcal P}_k(\bm{\theta})d^N\theta=
{\mathcal N}_k\int\limits_0^\pi \prod_{\sigma=1}^N
\exp\left[-\frac{(\theta_\sigma-f_{k\sigma})^2}{\phi_{k\sigma}^2}\right]
d^N\theta=1.
\edm

As we noted above, $P(\bm\theta)\not=0$ at $\left\{\theta_\sigma\in[0,\,\pi]\right\}$,
and therefore the distribution $P_2(\bm{\theta})$ as a sum of Gaussians is the most universal distribution,
because any other unequal to zero physical distribution can be represented in this form  with prescribed
precision.

From the correspondence rules for components of position and momentum operators in the optical tomography 
representation \cite{KorJRLR3274, KorPhysRevA85}
\be			\label{r46_1_18_1}
[\hat q_j]_w=
\sin\theta_j\partial_{X_j}^{-1}
\partial_{\theta_j}+X_j\cos\theta_j+
\frac{i\hbar}{2m_j\omega_{j}}\sin\theta_j\partial_{X_j},
\ee
\be			\label{r46_1_18_2}
[\hat p_j]_w=
m_j\omega_{j}\left(-\cos\theta_j\partial_{X_j}^{-1}
\partial_{\theta_j}+X_j\sin\theta_j\right)-\frac{i\hbar}{2}
\cos\theta_j\partial_{X_j},
\ee
with the help of general formula (\ref{eq8}) we can find the correspondence rules in the optical joint 
probability representation. To this end we note that each of expressions (\ref{r46_1_18_1}),
(\ref{r46_1_18_2}) contains only one term non-commuting with $P(\bm\theta)$.
This term include the differentiation  over the phase $\partial_{\theta_j}$, 
and we should calculate the following expression:
\be			\label{transfpartial0}
P(\bm\theta)\partial_{\theta_j}P^{-1}(\bm\theta)=-P^{-1}(\bm\theta)\left[\partial_{\theta_j}P(\bm\theta)\right]
+\partial_{\theta_j}.
\ee
Substituting here $P(\bm\theta)=P_2(\bm\theta)$, we obtain
\be			\label{transfpartial}
P_2(\bm\theta)\partial_{\theta_j}P_2^{-1}(\bm\theta)=2P_2^{-1}(\bm\theta)
\sum_{k=1}^KQ_k
\frac{\theta_j-f_{kj}}{\phi_{kj}^2}
{\mathcal P}_k(\bm{\theta})
+\partial_{\theta_j}.
\ee
Thus, we can write 
\be			\label{Qjointopt}
[\hat q_j]_{\widetilde w}=
\sin\theta_j\partial_{X_j}^{-1}\left(2P_2^{-1}(\bm\theta)
\sum_{k=1}^KQ_k
\frac{\theta_j-f_{kj}}{\phi_{kj}^2}
{\mathcal P}_k(\bm{\theta})
+
\partial_{\theta_j}\right)+X_j\cos\theta_j+
\frac{i\hbar}{2m_j\omega_{j}}\sin\theta_j\partial_{X_j},
\ee
\be			\label{Pjointopt}
[\hat p_j]_{\widetilde w}=
m_j\omega_{j}\left[-\cos\theta_j\partial_{X_j}^{-1}
\left(2P_2^{-1}(\bm\theta)
\sum_{k=1}^KQ_k
\frac{\theta_j-f_{kj}}{\phi_{kj}^2}
{\mathcal P}_k(\bm{\theta})
+
\partial_{\theta_j}
\right)+X_j\sin\theta_j\right]-\frac{i\hbar}{2}
\cos\theta_j\partial_{X_j}.
\ee

The dual symbols in the form of regular generalized functions in the joint probability
representation are obtained from the relevant dual symbols for the optical tomogram
found in \cite{KorPhysRevA85} by the division on the distribution function $P(\bm\theta)$.
For example, for the average value of position
\bdm
\langle\hat q_j\rangle=\int w_{\hat q_j}^{(d)}(\mbf X,\bm\theta)
w(\mbf X,\bm\theta,t)d^NXd^N\theta=
\int w_{\hat q_j}^{(d)}(\mbf X,\bm\theta)P^{-1}(\bm\theta)
\widetilde w(\mbf X,\bm\theta,t)d^NXd^N\theta.
\edm
Consequently
$\widetilde w_{\hat q_j}^{(d)}(\mbf X,\bm\theta)=w_{\hat q_j}^{(d)}(\mbf X,\bm\theta)P^{-1}(\bm\theta)$,
and so on:
\bdm
\begin{tabular}{lll}
$\widetilde w_{\hat q_j}^{(d)}(\mbf X,\bm\theta)=\ds{\frac{2}{\pi^NP(\bm\theta)}X_j\cos\theta_j},$&~~~~&
$\widetilde w_{\hat p_j}^{(d)}(\mbf X,\bm\theta)=\ds{\frac{2m_j\omega_j}{\pi^NP(\bm\theta)}X_j\sin\theta_j},$\\[3mm]
$\widetilde w_{\hat q_j^2}^{(d)}(\mbf X,\bm\theta)=\ds{\frac{X_j^2}{\pi^NP(\bm\theta)}(1+2\cos2\theta_j)},$&~~~~&
$\widetilde w_{\hat p_j^2}^{(d)}(\mbf X,\bm\theta)=\ds{\frac{X_j^2m_j^2\omega_j^2}{\pi^NP(\bm\theta)}(1-2\cos2\theta_j)},$
\end{tabular}
\edm
\bdm
\widetilde w_{\hat q_j\hat p_j}^{(d)}(\mbf X,\bm\theta)=\ds{\frac{2m_j\omega_j}{\pi^NP(\bm\theta)}X_j^2\sin2\theta_j
+\frac{i\hbar}{2}.}
\edm
These formulae are correct for any distribution $P(\bm\theta)$ unequal to zero including for $P_2(\bm\theta)$.

The evolution equation for the joint probability distribution $\widetilde w(\mbf X,\bm\theta,t)$ 
is easily obtained from the  evolution equation for the optical tomogram $w(\mbf X,\bm\theta,t)$
found in Refs. \cite{KorJRLR3274}, \cite{KorJRLR32338}
\be
\partial_t w(\mbf X,\bm\theta,t)=
\left\{\sum_{\sigma=1}^N\omega_{\sigma}
\left[\cos^2\theta_\sigma\partial_{\theta_\sigma}
-\frac{1}{2}\sin2\theta_\sigma\Big(1+X_\sigma\partial_{X_\sigma}\Big)
\right]
+\frac{2}{\hbar}\mbox{Im}~V\Big(
[\hat{\mbf q}]_w\,,t
\Big)\right\}
w(\mbf X,\bm\theta,t),
\label{eq_45_1}
\ee
where components of the operator $[\hat{\mbf q}]_w$ in the arguments of the potential $V$
are given by the expression (\ref{r46_1_18_1}).
With the help of (\ref{transfpartial}) we have
\begin{eqnarray}                             
\partial_t \widetilde w(\mbf X,\bm\theta,t)&=&
\Bigg\{\sum_{\sigma=1}^N\omega_{\sigma}
\Bigg[\cos^2\theta_\sigma
\left(
2P_3^{-1}(\bm\theta)
\sum_{k=1}^KQ_k
\frac{\theta_\sigma-f_{k\sigma}}{\phi_{k\sigma}^2}
{\mathcal P}_k(\bm{\theta})
+\partial_{\theta_\sigma}
\right)
\nonumber \\[3mm]
&&-\frac{1}{2}\sin2\theta_\sigma\Big(1+X_\sigma\partial_{X_\sigma}\Big)\Bigg]
+\frac{2}{\hbar}\mbox{Im}~V\Big(
[\hat{\mbf q}]_{\widetilde w}\,,t
\Big)\Bigg\}
\,\widetilde w(\mbf X,\bm\theta,t),
\label{eq_45_2}
\end{eqnarray}
where components of the operator $[\hat{\mbf q}]_{\widetilde w}$ are given by (\ref{Qjointopt}).

The equation of the stationary states in the joint probability representation is
also easily derived from the corresponding equation in the optical tomography representation 
found in Refs. \cite{KorJRLR3274}, \cite{KorJRLR32338}  
\begin{eqnarray}
E~w_E({\mbf X},{\bm\theta})&=&\Bigg[\sum_{\sigma=1}^Nm_\sigma\omega_\sigma^2\bigg\{\frac{\cos^2\theta_\sigma}{2}
\partial_{X_\sigma}^{-2}
\left(\partial_{\theta_\sigma}^2+1\right)
-\frac{X_\sigma}{2}\partial_{X_\sigma}^{-1}
\left(\cos^2\theta_\sigma+\sin2\theta_\sigma
\partial_{\theta_\sigma}\right)\nonumber \\[3mm]
&&+\frac{X_\sigma^2}{2}\sin^2\theta_\sigma-\frac{\hbar^2}{8m_\sigma^2\omega_\sigma^2}\cos^2\theta_\sigma
\partial_{X_\sigma}^2\bigg\}
+\mbox{Re}~V\Big([\hat{\mbf q}]_w\Big)\Bigg]
w_E({\mbf X},{\bm\theta}).
\label{StatOpt}
\end{eqnarray}
Contrary to (\ref{eq_45_1}), this equation has the items with the double differentiation 
$\partial_{\theta_\sigma}^2$ not commuting with $P(\bm\theta)$. Therefore, we  should find the following expression:
\be			\label{transfpartial2}
P(\bm\theta)\partial_{\theta_j}^2P^{-1}(\bm\theta)=
\partial_{\theta_\sigma}^2-2\frac{\partial_{\theta_\sigma}P(\bm\theta)}{P(\bm\theta)}\partial_{\theta_\sigma}
+2\left(\frac{\partial_{\theta_\sigma}P(\bm\theta)}{P(\bm\theta)}\right)^2
-\frac{\partial_{\theta_\sigma}^2P(\bm\theta)}{P(\bm\theta)}.
\ee
Using (\ref{transfpartial0}) and (\ref{transfpartial2}) from (\ref{StatOpt}) we can obtain
\begin{eqnarray}
&&\!\!\!\!\!\!\!\!\!\!\!\!\!\!\!\!\!\!
E~\widetilde w_E({\mbf X},{\bm\theta})=\Bigg[\sum_{\sigma=1}^Nm_\sigma\omega_\sigma^2\Bigg\{\frac{\cos^2\theta_\sigma}{2}
\partial_{X_\sigma}^{-2}
\left[\partial_{\theta_\sigma}^2-2\frac{\partial_{\theta_\sigma}P(\bm\theta)}{P(\bm\theta)}\partial_{\theta_\sigma}
+2\left(\frac{\partial_{\theta_\sigma}P(\bm\theta)}{P(\bm\theta)}\right)^2
-\frac{\partial_{\theta_\sigma}^2P(\bm\theta)}{P(\bm\theta)}
+1\right]\nonumber \\[3mm]
&&-\frac{X_\sigma}{2}\partial_{X_\sigma}^{-1}
\left[\cos^2\theta_\sigma+\sin2\theta_\sigma
\left(\partial_{\theta_\sigma}-\frac{\partial_{\theta_\sigma}P(\bm\theta)}{P(\bm\theta)}
\right)\right]
+\frac{X_\sigma^2}{2}\sin^2\theta_\sigma-\frac{\hbar^2}{8m_\sigma^2\omega_\sigma^2}\cos^2\theta_\sigma
\partial_{X_\sigma}^2\Bigg\} \nonumber \\[3mm]
&&+\mbox{Re}~V\Big([\hat{\mbf q}]_{\widetilde w}\Big)\Bigg]
\widetilde w_E({\mbf X},{\bm\theta}).
\label{StatJointOpt}
\end{eqnarray}
If the distribution $P(\bm\theta)$ has only one peak
\bdm
P(\bm\theta)=\mathcal{N}\prod_{\sigma=1}^N\exp\left[
-\frac{(\theta_\sigma-f_\sigma)^2}{\phi_\sigma^2}
\right],
\edm
then formula  (\ref{StatJointOpt}) is converted to 
\begin{eqnarray}
&&\!\!\!\!\!\!\!\!\!\!\!\!\!\!\!\!\!\!
E~\widetilde w_E({\mbf X},{\bm\theta})=\Bigg[\sum_{\sigma=1}^Nm_\sigma\omega_\sigma^2\Bigg\{\frac{\cos^2\theta_\sigma}{2}
\partial_{X_\sigma}^{-2}
\left[\partial_{\theta_\sigma}^2+4\frac{\theta_\sigma-f_\sigma}{\phi_\sigma^2}\partial_{\theta_\sigma}
+4\frac{(\theta_\sigma-f_\sigma)^2}{\phi_\sigma^4}+\frac{2}{\phi_\sigma^2}
+1\right]\nonumber \\[3mm]
&&-\frac{X_\sigma}{2}\partial_{X_\sigma}^{-1}
\left[\cos^2\theta_\sigma+\sin2\theta_\sigma
\left(\partial_{\theta_\sigma}+2\frac{\theta_\sigma-f_\sigma}{\phi_\sigma^2}
\right)\right]
+\frac{X_\sigma^2}{2}\sin^2\theta_\sigma-\frac{\hbar^2}{8m_\sigma^2\omega_\sigma^2}\cos^2\theta_\sigma
\partial_{X_\sigma}^2\Bigg\} \nonumber \\[3mm]
&&+\mbox{Re}~V\Big([\hat{\mbf q}]_{\widetilde w}\Big)\Bigg]
\widetilde w_E({\mbf X},{\bm\theta}).
\label{StatJointOpt1}
\end{eqnarray}

\section{Conclusion}
To summarize, we point out the main results of this work.
The tomographic formulation of quantum mechanics based, for example, on optical
tomographic probability distribution of quantum state, uses the 
conditional probability distribution depending on the local oscillator phase
parameter. We expanded this approach studying properties 
of the joint probability distribution \cite{BeautyInPhys, PhysScrT153}, where the parameter is 
considered as extra random variable.
We illustrated that the conventional quantum mechanics can be constructed in terms of such joint
probability distributions where the quantum states are described by the functions
dependent on random arguments and time, contrary to the tomographic representations,
in which the tomographic parameters are not random, and contrary to Wigner \cite{Wigner32}, Husimi \cite{Husimi40}, and
Glauber-Sudarshan \cite{Glauber63, Sudarshan63} representations, where the corresponding functions describing the states
are not probability distributions at all.
We presented the general formalism for symbols of operators in symplectic and optical
joint probability  representations.
Taking the Gaussian functions as the distributions of the tomographic parameters 
we found the correspondence rules for most interesting physical operators 
and derived the expressions of the dual symbols of operators in the form of singular 
and regular generalized functions.
Also we obtained evolution equations and stationary states equations
for symplectic and optical joint probability distributions.


\end{document}